\documentstyle[12pt]{article}
\title{Anomaly in Symplectic Integrator}
\author{Hiroto Kobayashi \\ \\ 
Department of Natural Science and Mathematics, \\
Chubu University, Kasugai 487-8501, Japan}
\date{}
\begin{document}
\maketitle
\begin{abstract}
Effective Liouville operators of the first- and the second-order
symplectic integrators are obtained for the one-dimensional
harmonic-oscillator system. The operators are defined only when the
time step is less than two. Absolute values of the coordinate and the
momentum monotonically increase for large time steps.
\end{abstract}

Symplectic integrators are widely used not only for the Hamiltonian
dynamics. Higher-order ones are also studied as the product of
exponential operators \cite{Suz,Yos} and are performed in various
areas. However, the upper limit of the time step where the scheme
works is not studied yet. Suzuki \cite{Con} studied the convergence of
decompositions of exponential operators in a Banach space. Bourbaki
treated only for the complete and normed Lie algebra. In the present
Letter, we discuss the convergence of the scheme for a specific
unbounded system. 

Let us consider the following first-order symplectic integrator for
the one-dimensional harmonic oscillator described by the Hamiltonian 
${\cal H}= (p^2 + q^2)/2 $:
\begin{equation}
\left\{
\begin{array}{l}
p' = p - x q , \\
q' = q + x p'. \end{array} \right. \label{eq:SI1}
\end{equation}
No anomalies seem to exist because the above discretization scheme is
defined for any time step $x$. However, the absolute values of $p$ and
$q$ monotonically increase for $x \ge 2$.

The above anomaly is explained by the divergence of the ''effective''
Liouville operator which corresponds to the symplectic integrator. The
discretization scheme Eq.~(\ref{eq:SI1}) is expressed by
\begin{displaymath}
\left( \begin{array}{c}
p' \\
q'     \end{array} \right) =
(1+xA)(1+xB) 
\left( \begin{array}{c}
p \\
q      \end{array} \right) =
\exp(xA) \exp(xB) 
\left( \begin{array}{c}
p \\
q      \end{array} \right) 
\end{displaymath}
with two matrices 
\begin{displaymath}
A = \left(\begin{array}{rr}
0 & 0 \\
1 & 0     \end{array} \right) , \quad
B = \left(\begin{array}{rr}
0 & -1 \\
0 &  0    \end{array} \right) 
\end{displaymath}
because of $A^2=0, \  B^2=0$. The equations 
$ABA=-A$ and $BAB=-B$ hold, and then we have
\begin{equation} 
\exp(xA) \exp(xB) = 
\exp \left\{ 
x F(x) \left(A + B + \frac{x}{\, 2 \,} [A,B] \right) 
\right\}, \label{eq:res1}
\end{equation}
where the function $F(x)$ is defined by
\begin{displaymath}
F(x) = \sum_{n=0}^\infty \frac{(n!)^2}{(2n+1)!} \,  x^{2n}
\end{displaymath}
with the convergence radius two \cite{epl}.

Defining an operator $L_1$ by
\begin{displaymath}
L_1 = A + B + \frac{x}{\, 2 \,} [A,B] 
= \left(\begin{array}{rr}
\frac{x}{\, 2 \,} &          -1 \\
       1          & - \frac{x}{\, 2 \,}    \end{array} \right) ,
\end{displaymath}
we have $i F(x) L_1$ as an effective Liouville operator which
corresponds to the symplectic integrator Eq.~(\ref{eq:SI1}). When 
a conserved quantity $E_1$ is expressed as a quadratic form of $p$ and
$q$ using a real symmetry matrix $M_1$ such as
\begin{displaymath}
E_1 = 
(p \  q) M_1
\left( \begin{array}{c}
p \\
q      \end{array} \right) ,
\end{displaymath}
the matrix $M_1 L_1$ is alternate, because the time derivative of $E_1$
is obtained from
\begin{eqnarray}
\frac{d E_1}{dt} & = & 
(p \  q) {}^t L_1 F(x) M_1
\left( \begin{array}{c}
p \\
q      \end{array} \right)  +
(p \  q) M_1 F(x) L_1
\left( \begin{array}{c}
p \\
q      \end{array} \right) \nonumber \\ & = & F(x) 
(p \  q) ({}^t (M_1 L_1) + M_1 L_1)
\left( \begin{array}{c}
p \\
q      \end{array} \right). \nonumber
\end{eqnarray}
Therefore, with
\begin{displaymath}
M_1 L_1 = \frac{1}{\, 2 \,} \det(L_1) 
\left(\begin{array}{rr}
0 & -1 \\
1 &  0    \end{array} \right) ,
\end{displaymath}
we have 
\begin{displaymath}
M_1 = \frac{1}{\, 2 \,}
\left(\begin{array}{rr}
0 & -1 \\
1 &  0    \end{array} \right) 
\left(\begin{array}{rr}
- \frac{x}{\, 2 \,} &        1 \\
         -1         & \frac{x}{\, 2 \,}    \end{array} \right) 
= \frac{1}{\, 2 \,}
\left(\begin{array}{rr}
          1         & - \frac{x}{\, 2 \,}  \\
- \frac{x}{\, 2 \,} &           1          \end{array} \right) ,
\end{displaymath}
that is, 
\begin{displaymath}
E_1 = \frac{1}{\, 2 \,} (p^2 - x p q + q^2).
\end{displaymath}

Within the convergence radius, the above Liouville operator 
$i F(x) L_1$ describes continuous dynamics along an ellipse in the 
$(p,q)$-plane with a constant value of $E_1$. Divergence of $F(x)$ can
be regarded as a break down of the continuous movement. 

Here we derive Eq.~(\ref{eq:res1}).

Expanding $\log(\exp(xA) \exp(xB))$ as the formal power series of 
$A$ and $B$, the following four kinds of terms appear:
\begin{eqnarray}
\underbrace{A B A \cdots B A}_{2n+1}, & & \nonumber \\
\underbrace{B A B \cdots A B}_{2n+1}, & & \nonumber \\
\underbrace{A B A \cdots A B}_{2n+2}, & & \nonumber \\
\underbrace{B A B \cdots B A}_{2n+2}  & & \nonumber
\end{eqnarray}
because of $A^2=0, \  B^2=0$. Coefficients of the above terms are
given by Goldberg's theorem \cite{Gld} as follows:
\begin{eqnarray}
 (-1)^n \frac{(n!)^2}{(2n+1)!}   \, x^{2n+1} ,&   & \nonumber \\
 (-1)^n \frac{(n!)^2}{(2n+1)!}   \, x^{2n+1} ,&   & \nonumber \\
 (-1)^n \frac{n!(n+1)!}{(2n+2)!} \, x^{2n+2} & = & 
 (-1)^n \frac{(n!)^2}{2(2n+1)!}  \, x^{2n+2} , \nonumber \\
-(-1)^n \frac{n!(n+1)!}{(2n+2)!} \, x^{2n+2} & = &
-(-1)^n \frac{(n!)^2}{2(2n+1)!}  \, x^{2n+2} . \nonumber
\end{eqnarray}
Because of $ABA=-A, \  BAB=-B$, we obtain 
\begin{eqnarray}
\underbrace{A B A \cdots B A}_{2n+1} & = & (-1)^n A  , \nonumber \\
\underbrace{B A B \cdots A B}_{2n+1} & = & (-1)^n B  , \nonumber \\
\underbrace{A B A \cdots A B}_{2n+2} & = & (-1)^n A B, \nonumber \\
\underbrace{B A B \cdots B A}_{2n+2} & = & (-1)^n B A, \nonumber
\end{eqnarray}
and we arrive at
\begin{eqnarray}
&   & \log(\exp(xA) \exp(xB))  \nonumber \\
& = & \sum_{n=0}^\infty \left\{
  \frac{(n!)^2}{(2n+1)!}   x^{2n+1} A + 
  \frac{(n!)^2}{(2n+1)!}   x^{2n+1} B +  
  \frac{(n!)^2}{2(2n+1)!}  x^{2n+2} A B 
- \frac{(n!)^2}{2(2n+1)!}  x^{2n+2} B A   \right\} \nonumber \\
& = & \sum_{n=0}^\infty \frac{(n!)^2}{(2n+1)!}  \,  x^{2n+1}  
\left(A + B + \frac{x}{\, 2 \,} [A,B] \right) \nonumber \\
& = & x F(x) \left(A + B + \frac{x}{\, 2 \,} [A,B] \right). \nonumber
\end{eqnarray}

Note that $\log(\exp(xA)\exp(xB))$ is expressed only with $A$, $B$,
and $[A,B]$ because of 
\begin{displaymath}
[A,[A,B]] = 2A, \quad [B,[A,B]] = - 2B.
\end{displaymath}

Next we consider the second-order symplectic integrator, which is
expressed with the two matrices $A$ and $B$ as
\begin{eqnarray}
\left( \begin{array}{c}
p' \\
q'     \end{array} \right) & = &
\left(1+\frac{x}{\, 2 \,}B \right) (1+xA) \left(1+\frac{x}{\, 2 \,}B \right) 
\left( \begin{array}{c}
p \\
q      \end{array} \right) \nonumber \\ & = &
\exp \left(\frac{x}{\, 2 \,}B \right) \exp(xA) \exp \left(\frac{x}{\, 2 \,}B \right) 
\left( \begin{array}{c}
p \\
q      \end{array} \right) . \label{eq:SI2}
\end{eqnarray}
Let us define $S(x)$ by
\begin{displaymath}
S(x) = \exp \left(\frac{x}{\, 2 \,}B \right) \exp(xA) 
\exp \left(\frac{x}{\, 2 \,}B \right) ,
\end{displaymath}
we obtain $S(-x) = S(x)^{-1}$ because of $S(x) S(-x) = 1$.
When we express $S(x)$ as
\begin{displaymath}
S(x) = \exp f(x),
\end{displaymath}
the following equation holds:
\begin{displaymath}
\exp f(-x) = S(-x) = S(x)^{-1} = \exp \{ -f(x) \},
\end{displaymath}
which indicates that $f(x)$ is an odd function. Therefore, when we
expand $f(x)$ as the formal power series of $A$ and $B$, the following
two kinds of terms appear:
\begin{eqnarray}
\underbrace{A B A \cdots B A}_{2n+1}, & & \label{eq:A} \\
\underbrace{B A B \cdots A B}_{2n+1}. & & \label{eq:B} 
\end{eqnarray}
Because of $ABA=-A, \  BAB=-B$, we can express $f(x)$ as
\begin{displaymath}
f(x) = x ( F_1 (x) A + F_2 (x) B )
\end{displaymath}
with two even functions $F_1 (x)$ and $F_2 (x)$. 

When we expand $\log (\exp X_1 \exp X_2 \exp X_3) $ as the formal
power series of $X_1$, $X_2$, and $X_3$, coefficients of the terms are
given by an extended version of Goldberg's theorem \cite{Reu,KHS}.
Because we later replace $X_1$, $X_2$, and $X_3$ as
\begin{eqnarray}
X_1 = X_3 & = & \frac{x}{\, 2 \,}B, \nonumber \\
X_2 & = & xA, \nonumber
\end{eqnarray}
two kinds of terms
\begin{eqnarray}
& & \underbrace{X_2 X_{i_1} X_2 X_{i_2} X_2 X_{i_3} X_2 
\cdots X_2 X_{i_n} X_2}_{2n+1}, \label{eq:X2} \\
& & \underbrace{X_{i_1} X_2 X_{i_2} X_2 X_{i_3} X_2 
\cdots X_2 X_{i_n} X_2 X_{i_{n+1}} }_{2n+1}  \label{eq:Xi}
\end{eqnarray}
contribute to the terms (\ref{eq:A}) and (\ref{eq:B}), respectively,
with $i_j = 1$ or $3$. According to Goldberg's theorem, coefficients
do not depend whether $X_{i_j}$ between two $X_2$'s is $X_1$ or $X_3$.
Then the coefficient of the term (\ref{eq:X2}) is given by
\begin{displaymath}
(-1)^n \frac{(n!)^2}{(2n+1)!}
\end{displaymath}
for all $X_{i_j}$. The coefficient of the term (\ref{eq:Xi}) is also
given by
\begin{eqnarray}
(-1)^n \frac{(n!)^2}{(2n+1)!} & {\rm for} & 
(i_1,i_{n+1})=(1,1), \  (3,3), \nonumber \\
(-1)^{n+1} \frac{(n-1)!(n+1)!}{(2n+1)!} & {\rm for} & 
(i_1,i_{n+1})=(1,3), \  (3,1), \nonumber
\end{eqnarray}
if $n \ge 1$. Replacing $X_1$, $X_2$, and $X_3$ with $A$ and $B$, and
using $ABA=-A$ and $BAB=-B$, we arrive at
\begin{eqnarray}
F_1 (x) & = & \sum_{n=0}^\infty \frac{(n!)^2}{(2n+1)!} \, x^{2n} =
F(x), \nonumber \\
F_2 (x) & = & 1 - \sum_{n=1}^\infty \frac{(n-1)!n!}{2(2n+1)!} \, x^{2n}
          = \left( 1 - \frac{x^2}{\,  4  \, } \right) F(x), \nonumber
\end{eqnarray}
that is,
\begin{displaymath}
\exp \left(\frac{x}{\, 2 \,}B \right) \exp(xA) 
\exp \left(\frac{x}{\, 2 \,}B \right) =
\exp \left\{ x F(x) \left(
A + \left( 1 - \frac{x^2}{\,  4  \, } \right) B \right) \right\} .
\end{displaymath}

As for the first-order scheme, defining $ L_2 $ by
\begin{displaymath}
L_2 = A + \left( 1 - \frac{x^2}{\,  4  \, } \right) B,
\end{displaymath}
we can regard $i F(x) L_2$ as an effective Liouville operator which
corresponds to the second-order symplectic integrator
Eq.~(\ref{eq:SI2}). Then we obtain the conserved quantity $E_2$ as
\begin{displaymath}
E_2 = \frac{1}{\, 2 \,} \left( p^2 + 
               \left( 1 - \frac{x^2}{\,  4  \, } \right) q^2 \right).
\end{displaymath}

Finally we note that the exponential function absolutely converges and
the product of exponentials always exists, which corresponds to the
fact that the discretization scheme as Eqs.~(\ref{eq:SI1}) and
(\ref{eq:SI2}) can be defined for all time steps. However, the
absolute values of $p$ and $q$ monotonically increase for large time
steps although Jacobian of the time-evolution operator is unity. 

I thank Profs.\ M. Suzuki and N. Hatano for their useful comments and
suggestions. I am also grateful to Dr.\ H. Watanabe for fruitful
discussions.


\begin{thebibliography}{9}
\bibitem{epl} The function $F(x)$ can be given in a compact form as  
$ 2 \arctan (x / 2 \sqrt{1-x^2/4} )$ $/ x \sqrt{1-x^2/4} $ for $0<|x|<2$.
\end{thebibliography}
\end{document}